\documentstyle[12pt]{article}
\begin{document}
\thispagestyle{empty}

\begin{center}
\LARGE \tt \bf{On the existence of Kalb-Ramond wormholes}
\end{center}

\vspace{2.5cm}

\begin{center} {\large By  L.C. Garcia de Andrade\footnote{Departamento de
F\'{\i}sica Te\'{o}rica - Instituto de F\'{\i}sica - UERJ

Rua S\~{a}o Fco. Xavier 524, Rio de Janeiro, RJ

Maracan\~{a}, CEP:20550-003 , Brasil.}}
\end{center}

\vspace{2.0cm}

\begin{abstract}
Static spherically symmetric Kalb-Ramond (KR) traversable wormholes in torsioned spacetime are shown not to exist globally since torsion vanishes not only at the wormhole throat but in all spacetime points which means that all components of the KR field vanish identically. With this result Hochberg-Visser result on the non-violation of the weak energy (WEC) condition in the presence of four-dimensional Einstein-Cartan-Kalb-Ramond (ECKR) geometry with totally skew-symmetric Cartan torsion seems to be a trivial one.
\end{abstract}

\newpage
\pagestyle{myheadings}
\markright{\underline{ECKR wormholes?}}
\section{Introduction}

Recently Hochberg and Visser \cite{1} have investigated in detail the null energy conditions for wormholes \cite{2} in the totally skew symmetric torsioned spacetime where torsion is proportional to Kalb-Ramond fields of closed strings of low-energy physics.They showed that the presence of KR fields insteat of allowing the energy condition violations be dumped into torsion degrees of freedom they worsen the WEC and do not collaborate to the wormhole throat stability but helps to collapse it. In this paper we show that in fact this analysis would not be necessary since static spherically symmetric KR wormholes are forbidden in torsioned spacetime since torsion vanishes not only locally in the wormhole throat but actually torsion vanishes globally in wormhole spacetime. This simple proof is obtained by making use  based on the recently obtained result by SenGupta and his group \cite{3} which show that static spherically symmetric Kalb-Ramond geometry like Scharzschild metric can be obtained. Actually one can say that wormholes of anytype are forbidden in KR torsioned spacetimes since as pointed out recently by SenGupta and myself \cite{4,5} there are no time dependent solutions of ECKR gravity field equations. The proof is straightforward and comes from the ECKR system \cite{3}
\begin{equation}
e^{-{\lambda}}(\frac{1}{r^{2}}-\frac{{\lambda}'}{r})-\frac{1}{r^{2}}=k(h_{1}h^{1}+h_{2}h^{2}+h_{3}h^{3}-h_{4}h^{4})
\label{1}
\end{equation}

\begin{equation}
e^{-{\lambda}}(\frac{1}{r^{2}}-\frac{{\nu}'}{r})-\frac{1}{r^{2}}=k(h_{1}h^{1}+h_{2}h^{2}-h_{3}h^{3}+h_{4}h^{4})
\label{2}
\end{equation}

\begin{equation}
e^{-{\lambda}}({\nu}" +\frac{{{\nu}'}^{2}}{2}-\frac{{\nu}'{\lambda}'}{2}+\frac{{\nu}'-{\lambda}'}{r})-e^{-{\nu}}(\ddot{\lambda}+\frac{{\dot{\lambda}}^{2}}{2}-\frac{\dot{\lambda}\dot{\nu}}{2})=2k(h_{1}h^{1}-h_{2}h^{2}+h_{3}h^{3}+h_{4}h^{4})
\label{3}
\end{equation}

\begin{equation}
e^{-{\lambda}}({\nu}" +\frac{{{\nu}'}^{2}}{2}-\frac{{\nu}'{\lambda}'}{2}+\frac{{\nu}'-{\lambda}'}{r})-e^{-{\nu}}(\ddot{\lambda}+\frac{{\dot{\lambda}}^{2}}{2}-\frac{\dot{\lambda}\dot{\nu}}{2})=2k(-h_{1}h^{1}+h_{2}h^{2}+h_{3}h^{3}+h_{4}h^{4})
\label{4}
\end{equation}
where we use the spherically symmetric geometry
\begin{equation}
ds^{2}=e^{{\nu}(r,t)}dt^{2}-e^{{\lambda}(r,t)}dr^{2}-r^{2}(d{\theta}^{2}+sin^{2}{\theta}{d{\phi}}^{2})
\label{5}
\end{equation}
and the remaining conditions on the ECKR gravity field are
\begin{equation}
e^{\lambda}\frac{\lambda}{r}= 2kh_{3}h^{4}
\label{6}
\end{equation}
\begin{equation}
h_{2}h^{4}=h_{1}h^{4}=h_{2}h^{3}=h_{1}h^{2}=h_{3}h^{1}=0
\label{7}
\end{equation}
Here $ H_{{\mu}{\nu}{\eta}}={\partial}_{[{\mu}}B_{{\nu}{\eta}]}$ is the strenght of the KR field $B_{{\mu}{\nu}}$. The KR field strenght components ${H_{012},H_{013},H_{023}}$ and $H_{123}$ are respectively denoted by $h_{1}$,$h_{2}$,$h_{3}$ and $h_{4}$. Since the LHS of equations (\ref{3}) and (\ref{4}) are identical from expressions (\ref{7}) one obtains
\begin{equation}
h_{1}=h_{2}=0
\label{8}
\end{equation}
with this great simplification the KR field equations \cite{3} read
\begin{equation}
{{h^{3}},}_{3}={{h^{4}},}_{4}=0	
\label{9}
\end{equation}
which means that both KR fields are not dependent of the coordinate ${\phi}$ and finally the remaining KR field equations are
\begin{equation}
{{h^{3}},}_{2}+cot{\theta}h^{3}=0
\label{10}
\end{equation}
and
\begin{equation}
{{h^{4}},}_{1}+(\frac{{\nu}'+{\lambda}'}{2} +\frac{2}{r})h^{4}=0
\label{11}
\end{equation}
and
\begin{equation}
{{h^{4}},}_{2}+cot{\theta}h^{4}=0
\label{12}
\end{equation}
The purely space dependent metric imply that the metric coefficients ${\nu}$ and ${\lambda}$ depend on the radial coordinate r. We consider here just the case where $h_{4}=0$ and only $h_{3}$ component survives since the converse has been proved by SenGupta and Sur \cite{3} to be unphysical in the sense that torsion would give a finite value as the radial coordinate r approaches infinity.Therefore as shown by SenGupta and Sur \cite{3} the solution is given by
\begin{equation}
e^{\lambda}={(1+\frac{c_{1}}{r}+\frac{{\tau}{r}}{r})}^{-1}
\label{13}
\end{equation}
and 
\begin{equation}
e^{\nu}=\frac{c_{2}}{r(r+{\tau}+c_{1})}exp[\int{\frac{2dr}{r+{\tau}+c_{1}}}]
\label{14}
\end{equation}
To apply the SenGupta-Sur solution to a static wormhole let us consider a general wormhole spacetime \cite{2} line element given by 
\begin{equation}
ds^{2}= e^{2{\phi}(r,t)}dt^{2}-\frac{dr^{2}}{(1-\frac{b(r)}{r})}-r^{2}(d{\theta}^{2}+sin^{2}{\theta}{d{\phi}}^{2})
\label{15}
\end{equation}
This is the Morris-Thorne wormhole \cite{6}, where $b(r)$ is the wormhole throat parameter which vanishes when the wormhole throat collapses and 
\begin{equation}
{\tau}(r)= k\int{r^{2}h^{2}(r)dr}
\label{16}
\end{equation} 
Since as shown by SenGupta and Sur \cite{3} $h(r)=\frac{\beta}{r^{2}}$ by comparing the solutions (\ref{13}) and (\ref{15}) one obtains  
\begin{equation}
e^{-{\lambda}}=(1+\frac{c_{1}}{r}-\frac{\beta}{r^{2}})
\label{17}
\end{equation}
and 
\begin{equation}
-c_{1}r+{\beta}={\beta}(r)
\label{18}
\end{equation}
where we have computing ${\tau}$ by from the integral (\ref{16}) and the value of $h^{2}$ yielding ${\tau}=-\frac{{\beta}}{r}$.Here $c_{1}$ and $c_{2}$ are integration constants. 
By applying the traversable wormhole condition
\begin{equation}
b(r_{0})=r_{0}
\label{19}
\end{equation}
at the wormhole throat into the condition (\ref{18}) one obtains
\begin{equation}
b(r_{0})=-c_{1}r_{0}+{\beta}
\label{20}
\end{equation}
which  yields the following conditions
\begin{equation}
c_{1}=-1
\label{21}
\end{equation}
and
\begin{equation}
{\beta}= 0
\label{22}
\end{equation}
Since ${\beta}$ is a constant this result is not only a local result valid only at the wormhole throat but is actually a global result valid thoughout spacetime. Of course this also means that the only nonvanishing KR field strenght component $h_{3}$ vanishes. These last two constraints on the solution yields do not means that there are no torsioned wormholes at all but that there are no KR wormholes at all but that there are no spherically symmetric KR wormholes time dependent or not. Actually cosmological torsioned KR wormholes or Kerr wormholes could be found in ECKR theory of gravity. This investigation is now under progress.
\section*{Acknowledgement}
I am very much indebt to Prof. Soumitra SenGupta ,Prof. K.S. Thorne and Prof. Matt Visser for helpful discussions on the subject of this  paper. Financial support from CNPq. and UERJ are gratefully ackowledged.


\begin{thebibliography}{6}
\bibitem{1} D. Hochberg and M. Visser,Phys. Rev.D 58 (1998) 044021.
\bibitem{2} S. SenGupta and S.Sur,Phys. Lett. B514 (2001)109.
\bibitem{3} M. Visser, Lorentzian Wormholes:From Einstein to Hawking (1996) AIP. 
\bibitem{4} S.SenGupta,private communication.
\bibitem{5} L.C. Garcia de Andrade, A comment on the inhomogeneous Einstein-Cartan-Kalb-Ramond fields in Cosmology,gr-qc/0203089.
\bibitem{6} M. Morris and K.S.Thorne, Am. J. Phys. 56(1988) 395.
\end{thebibliography}
\end{document}